# The influence of vibronic coupling on the shape of transport characteristics in inelastic tunneling through molecules

Kamil Walczak

Institute of Physics, Adam Mickiewicz University
Umultowska 85, 61-614 Poznań, Poland

Here we present theoretical studies of the effect of vibronic coupling on nonlinear transport characteristics (current-voltage and conductance-voltage) in molecular electronic devices. Considered device is composed of molecular quantum dot (with discrete energy levels) weakly connected to metallic electrodes (treated within the wide-band approximation), where molecular vibrations are modeled as dispersionless phonon excitations. Nonperturbative computational scheme, used in this work, is based on the Green's function theory within the framework of mapping technique (GFT-MT) which transforms the many-body electron-phonon interaction problem into a one-body multi-channel single-electron scattering problem. In particular, it is shown that quantum coherent transport of virtual polarons through the molecule can be a dominant factor justifying some well-known discrepancies between theoretical calculations and experimental results.

## 1. Introduction

Inelastic electron tunneling spectroscopy (IETS) is a powerful experimental tool for identifying and characterizing molecular species within the conduction region [1-17]. Standard ac modulation techniques, along with two lock-in amplifiers, are utilized to measure current-voltage ($I-V$) characteristics as well as the first and second harmonic signals (proportional to $dI/dV$ and $d^2I/dV^2$, respectively). This method provides information on the strength of the vibronic coupling between the charge carriers and nuclear motions of the molecules. The IETS experiment can also be helpful in identifying the geometrical structures of molecules and molecule-metal contacts, since junctions with different geometries disclose very different spectral profiles [16,17]. The measured spectra show well-resolved vibronic features corresponding to certain vibrational normal modes of the molecules. It is also well-known that the IETS spectra are very sensitive also to few other factors, such as: (i) the device working temperature, (ii) the strength of the molecule-metal bonding, and (iii) the intramolecular conformational changes.

In the literature, we can distinguish two different approaches to the problem of inelastic transport. One of them is based on low order perturbative treatment, where the tunneling current is computed in the lowest order in the electron-phonon coupling [18-22]. However, this approach is not fully consistent with the nonequilibrium conditions under which such measurements are done as well as with the boundary restrictions imposed by the Pauli principle. The second mentioned method is associated with non-perturbative treatment, where the many-body electron-phonon interaction problem is transformed into a one-body many-channel scattering problem within the so-called mapping technique [23-32]. It is believed that this approximation is well-justified in the boundary case of the high voltages, while it does not involve any restrictions on the model parameters.

The main purpose of this work is to use non-perturbative method, based on Green's function theory and mapping technique (GFT-MT) [10-16], to study the effect of vibronic coupling on the shape of transport characteristics (current-voltage and conductance-voltage) in the IETS experiments. We hypothetically divide the molecular device into three parts, where the central molecular bridge is isolated from two electrodes (source and drain) via potential barriers. Since the molecule is weakly connected to the electrodes, a molecule itself is treated as a quantum dot with discrete energy levels, while source and drain are described within a broad-band theory. Molecular vibrations are modelled as dispersionless (Einstein) phonon excitations which can locally interact with conduction electrons. The electrons passing through energetically accessible molecular states (conducting channels) may exchange a definite amount of energy with the nuclear degrees of freedom, resulting in an inelastic component to the electrical current. Such molecular oscillations can have essential influence on the shape of transport characteristics especially in the case, when the residence time of a tunneling electron on a molecular bridge is of order of magnitude of the time involved in nuclear vibrations ($\sim ps$).

## 2. Theory

Now we briefly outline our theoretical approach. Let us write the full Hamiltonian of considered system as a sum:

$$H = \sum_{\alpha} H_{\alpha} + H_M + H_T, \quad (1)$$

where: $\alpha = L$ for left (source) electrode and $\alpha = R$ for right (drain) electrode, respectively, in the case of two-terminal junction. Both metallic electrodes are treated as reservoirs for non-interacting electrons and described with the help of the following Hamiltonian:

$$H_L + H_R = \sum_{k \in \alpha} \varepsilon_k c_k^+ c_k. \quad (2)$$

Here: $\varepsilon_k$ is the single particle energy of conduction electrons, while $c_k^+$ and $c_k$ denote the electron creation and annihilation operators with momentum $k$ in the $\alpha$ electrode. The third term describes molecular bridge with Holstein-type phonons:

$$H_M = \sum_{i,j} [\varepsilon_i - \lambda_j (a_j + a_j^+)] d_i^+ d_i + \sum_j \Omega_j a_j^+ a_j, \quad (3)$$

Here: $\varepsilon_i$ is single energy level of molecular quantum dot, $\Omega_j$ is phonon energy in the $j$ mode, $\lambda_j$ is the strength of on-level electron-phonon interaction. Furthermore, $d_i^+$ and $d_i$ are electron creation and annihilation operators on level $i$, while $a_j^+$ and, $a_j$ are phonon creation and annihilation operators, respectively. The last term represents the coupling of molecular quantum dot to the electrodes:

$$H_T = \sum_{k \in \alpha; i} [\gamma_{k,i} c_k^+ d_i + h.c.], \quad (4)$$

where the matrix elements $\gamma_{k,i}$ stands for the strength of the tunnel coupling between the dot and metallic electrodes.

The problem we are facing now is to solve a many-body problem with phonon emission and absorption when the electron tunnels through the dot. Let us consider for transparency only one phonon mode (primary mode), since generalization to multi-phonon case can be obtained straightforwardly. The electron states into the dot are expanded onto the direct product states composed of single-electron states and $m$-phonon Fock states:

$$|i,m\rangle = d_i^+ \frac{(a^+)^m}{\sqrt{m!}}|0\rangle, \tag{5}$$

where electron state $|i\rangle$ is accompanied by $m$ phonons ($|0\rangle$ denotes the vacuum state). Similarly the electron states in the electrodes can be expanded onto the states:

$$|k,m\rangle = c_k^+ \frac{(a^+)^m}{\sqrt{m!}}|0\rangle, \tag{6}$$

where the state $|k\rangle$ with momentum $k$ is accompanied by $m$ phonons. In this procedure, the non-interacting single-mode electrodes (Eq.2) are mapped to a multi-channel model:

$$\tilde{H}_L + \tilde{H}_R = \sum_{k \in \alpha; m}(\varepsilon_{k\sigma} + m\Omega)|k,m\rangle\langle k,m|. \tag{7}$$

Since the channel index $m$ represents the phonon quanta excited in the reservoir, accessibility of particular conduction channels is determined by a weight factor:

$$P_m = [1 - \exp(-\beta\Omega)]\exp(-m\beta\Omega). \tag{8}$$

Here Boltzmann distribution function is used to indicate the statistical probability of the phonon number state $|m\rangle$ at finite temperature $\theta$, $\beta^{-1} = k_B\theta$ and $k_B$ is Boltzmann constant. Since we neglect nonequilibrium phonon effects and dissipative processes, the system conserves its total energy during the scattering process and therefore the electron energies are constrained by the following energy conservation law:

$$\varepsilon_{in} + m\Omega = \varepsilon_{out} + n\Omega. \tag{9}$$

Moreover, in practice, the basis set is truncated to a finite number of possible excitations $m = m_{max}$ in the phonon modes because of the numerical efficiency. The size of the basis set strongly depends on: (i) phonon energy, (ii) temperature in the reservoir under investigation and (iii) the strength of the electron-phonon coupling constant. In the new representation (Eq.5), molecular Hamiltonian (Eq.3) can be rewritten in the form:

$$\tilde{H}_M = \sum_{i,m}(\varepsilon_i + m\Omega)|i,m\rangle\langle i,m| - \sum_{i,m}\lambda\sqrt{m+1}(|i,m+1\rangle\langle i,m| + |i,m\rangle\langle i,m+1|), \tag{10}$$

which for each molecular energy level $i$ is analogous to tight-binding model with different site energies and site-to-site hopping integrals. Finally, the tunneling part can also be rewritten in terms of considered basis set as:

$$\tilde{H}_T = \sum_{k \in \alpha; i,m}(\gamma_{k,i}^m|k,m\rangle\langle i,m| + h.c.), \tag{11}$$

where $\gamma_{k,i}^{m}$ is the coupling between the $m$ th pseudochannel in the electrode and the molecular system, respectively. To avoid unnecessary complexities, in further analysis we take into account molecular bridge which is represented by one electronic level – generalization to multilevel system is simple.

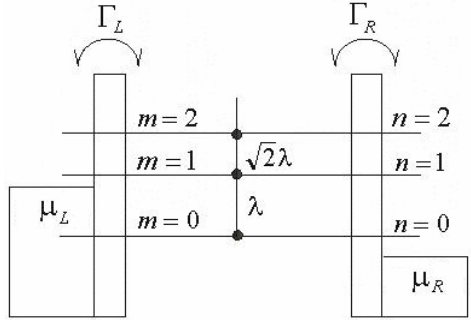

Figure 1: A schematic representation of inelastic scattering problem for the device composed of molecular quantum dot with single energy level connected to two reservoirs.

Now we proceed to analyze the problem of electron transfer between two reservoirs of charge carriers via molecular quantum dot in the presence of phonons. An electron entering from the left hand side can suffer inelastic collisions by absorbing or emitting phonons before entering the right electrode. Such processes are presented graphically in Fig.1, where individual channels are indexed by the number of phonon quanta in the left $m$ and right electrode $n$, respectively. Each of the possible processes is described by its own transmission probability, which can be written in the factorized form:

$$T_{m,n}(\varepsilon) = \Gamma_L \Gamma_R \left| G_{m+1,n+1}(\varepsilon) \right|^2. \qquad (12)$$

Such transmission function (Eq.12) is expressed in terms of the so-called linewidth functions $\Gamma_\alpha$ ($\alpha = L, R$) and the matrix element of the Green's function defined as:

$$G(\varepsilon) = \left[ 1\varepsilon - \tilde{H}_M - \Sigma_L - \Sigma_R \right]^{-1}. \qquad (13)$$

Here: 1 stands for identity matrix, $\tilde{H}_M$ is the molecular Hamiltonian (Eq.10), while the effect of the electronic coupling to the electrodes is fully described by specifying self-energy corrections $\Sigma_\alpha$.

In the present paper we adopt wide-band (WB) approximation to treat metallic electrodes, where the hopping matrix element is independent of energy and bias voltage, i.e. $\gamma_{k,i}^{m} = \gamma_\alpha$. In this case, the self-energy is given through the relation:

$$\Sigma_\alpha = -\frac{i}{2}\Gamma_\alpha, \qquad (14)$$

where

$$\Gamma_\alpha = 2\pi \left| \gamma_\alpha \right|^2 \rho_\alpha. \qquad (15)$$

Here: $\rho_\alpha$ is density of states in the $\alpha$-electrode. This self-energy function is mainly responsible for level broadening and generally depends on: (i) the material that the electrode is made of, and (ii) the strength of the coupling with the electrode. There are few factors that can be crucial in determining the parameter of the coupling strength, such as: (i) the atomic-scale contact geometry, (ii) the nature of the molecule-to-electrode coupling (chemisorption or physisorption), (iii) the molecule-to-electrode distance, or even (iv) the variation of the surface properties due to adsorption of molecular monolayer. The consequences of WB approximation are: (i) negligence of the resonance shift due to the coupling with the electrodes, (ii) the loss of the correct description of the contact and (iii) quantitative error of order 30 % in the magnitude of calculated current [33]. However, our essential conclusions can be generalized well beyond this simplification. Both electrodes are also identified with their electrochemical potentials [34]:

$$\mu_L = \varepsilon_F - \eta eV \tag{16}$$

and

$$\mu_R = \varepsilon_F + (1-\eta)eV, \tag{17}$$

which are related to the Fermi energy level $\varepsilon_F$. The voltage division factor $0 \leq \eta \leq 1$ describes how the electrostatic potential difference $V$ is divided between two contacts and can be related to the relative strength of the coupling with two electrodes: $\eta = 2^{-\gamma_L/\gamma_R}$. In our analyses we can distinguish two boundary cases: $\gamma_L = \gamma_R \Rightarrow \eta = 1/2$ for interpretation of mechanically controllable break-junctions (MCB experiments) and $\gamma_L \gg \gamma_R \Rightarrow \eta \approx 0$ for interpretation of scanning tunneling microscopy (STM experiments), respectively. Here we assume the symmetric coupling case, where $\eta = 1/2$. It should also be noted that the case of asymmetric coupling ($\eta \neq 1/2$) generates rectification effect [35].

The total current flowing through the junction can be expressed in terms of transmission probability of the individual transitions $T_{m,n}$ which connects incoming channel $m$ with outgoing channel $n$:

$$I_{tot}(V) = \frac{2e}{h} \int_{-\infty}^{+\infty} d\varepsilon \sum_{m,n} T_{m,n} \left[ P_m f_L^m (1 - f_R^n) - P_n f_R^n (1 - f_L^m) \right], \tag{18}$$

where:

$$f_\alpha^m = \left[ \exp[\beta(\varepsilon + m\omega - \mu_\alpha)] + 1 \right]^{-1} \tag{19}$$

is the equilibrium Fermi distribution function. The factor of 2 in Eq.18 accounts for the two spin orientations of conduction electrons. The elastic contribution to the current is obtained can be obtained from Eq.18 by imposing the constraint of elastic transitions, where $\varepsilon_{in} = \varepsilon_{out}$ or more precisely $m = n$:

$$I_{el}(V) = \frac{2e}{h} \int_{-\infty}^{+\infty} d\varepsilon \sum_m T_{m,m} P_m \left[ f_L^m - f_R^m \right]. \tag{20}$$

The differential conductance is then given by the derivative of the current with respect to voltage: $G(V) = dI(V)/dV$.

## 3. Results and discussion

Now we proceed to consider one electronic level $\varepsilon_0$ which is connected to two broad-band paramagnetic electrodes, where the electrons on the dot are coupled with the coupling strength $\lambda$ to a single phonon mode with energy $\Omega$ (primary mode). This is a test case simple enough to analyze the essential physics of inelastic transport problem in detail. Besides, generalization to multilevel system with many different phonons can be obtained straightforwardly. All of the key parameters can be inferred from different experimental data. The width parameters $\Gamma_L$ and $\Gamma_R$ are related to observed lifetimes of excess electrons on molecules adsorbed on metal surfaces and can be estimated from time-resolved two-photon photoemission experiments to be in the range of 0.1-1 eV for chemisorbed species [36,37]. The electron-phonon coupling parameter $\lambda$ can be estimated in molecular systems from reorganization energies: $E_{reorg} \approx \lambda^2/\Omega$, inferred from electron-transfer rate studies in similar environments. Since observed values of $E_{reorg}$ are 0.1-1 eV and $\Omega \sim 0.1$ eV, the magnitude of $\lambda$ is placed in the range of 0.1-0.3 eV. Finally, we choose the maximum number of the allowed phonon quanta $m_{max} = 8$ to obtain converged results for all the parameters involved in this paper.

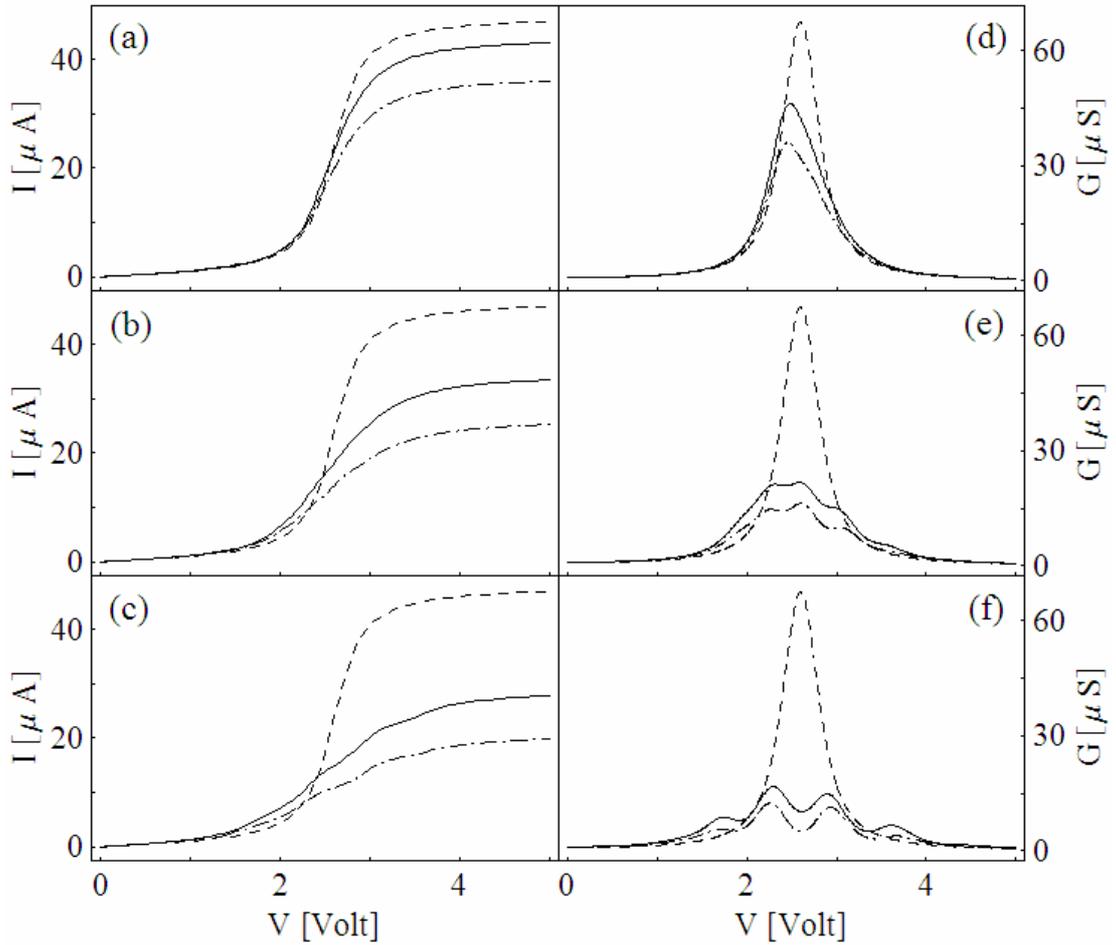

Figure 2: The current-voltage $I(V) = -I(-V)$ (a, b, c) and conductance-voltage $G(V) = G(-V)$ (d, e, f) characteristics for three different values of vibronic coupling parameters: $\lambda = 0.1$ (a, d), $\lambda = 0.2$ (b, e), and $\lambda = 0.3$ (c, f). Total current (solid line) and its elastic part (dashed-dotted line) are compared with the current obtained in the absence of phonons (dashed line). The other parameters of the model (given in eV): $\varepsilon_0 = 0$, $\varepsilon_F = -1.3$, $\Omega = 0.13$, $\Gamma_L = \Gamma_R = 0.1$, $\beta^{-1} = 0.025$.

The dependence of the electrical current on bias voltage for three different values of the $\lambda$-parameter is illustrated in Figs.2a-c. Since molecular vibrations are observed with equal intensity in the positive and negative bias polarity, thus for clarity we only show the positive bias region in the spectrum. Here we can observe the general tendency: the stronger the vibronic coupling, the smoother the $I-V$ characteristic and the lower values of the current flowing through the junction. The conductance-voltage ($G-V$) functions for three different values of the $\lambda$-parameter are demonstrated in Figs.2d-f. Here we can see that an amount of the visible peaks strongly depends on the vibronic coupling. Generally, the stronger the $\lambda$-parameter, the more peaks involved and the longer voltage distance between them. It can be deduced from the formula for polaron energies that the positions of the peaks in the conductance spectrum approximately coincide with the following expression:

$$V \approx 2\left[V_0 - \frac{\lambda^2}{e\Omega} + \frac{m\Omega}{e}\right], \quad (21)$$

where $V_0 = |\varepsilon_F - \varepsilon_0|/e$ corresponds to the bias voltage for the case of non-phonon conductance peak. On the other hand, for the small values of $\lambda$, we can not distinguish particular peaks since they are merged into one misshapen peak. Because of this overlapping it could be difficult to determine the intrinsic line width of a single vibration mode.

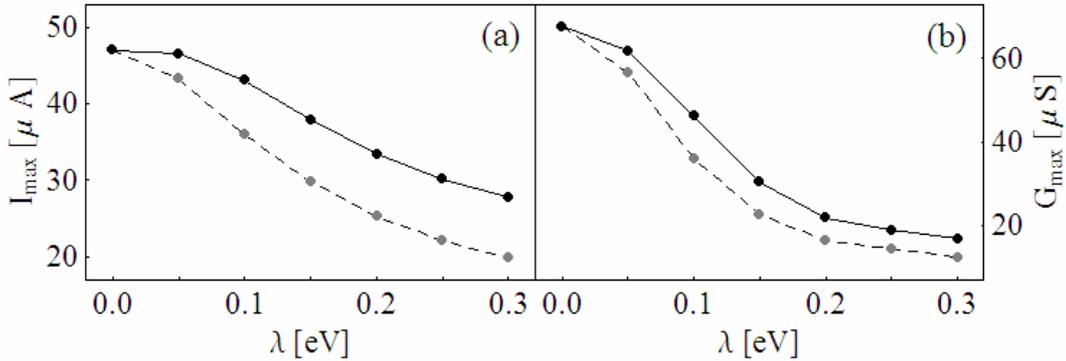

Figure 3: Maximum value of the current $I_{max}$ (a) and maximal height of the conductance peak $G_{max}$ (b) are plotted against the vibronic coupling strength $\lambda$. Total values of the current (black circles, solid line) are compared with its elastic part (grey circles, dashed line). The other parameters of the model are the same as in Fig.2.

Furthermore, one of the most essential controversies in molecular electronics is associated with the fact that the calculated currents for single-molecule devices is usually two or even three orders of magnitude overestimated in comparison with experimental data [38]. In Fig.3a we plot the maximal value of the current $I_{max}$ (calculated for $V = 5$ Volts) as a function of the vibronic coupling $\lambda$. An increase of the $\lambda$-parameter from 0 to 0.3 eV results in reduction of the magnitude of the current flowing through the junction of about 40 %, while its elastic contribution is suppressed of about 60 %. At the same time, the maximal height of the conductance peak is reduced of about 75 %, while the suppression of its elastic contribution is over 80 %, as documented in Fig.3b. So far, this divergence was closely connected to some coupling effects, such as: (i) the atomic-scale contact geometry, (ii) the nature of molecule-metal coupling, or even (iii) the changes of surface properties due to adsorption of molecular layers (during the preparation of the sample). However, here we show that also the effect of inelastic tunneling due to polaron formation can reduce the current at the molecular scale. Besides, it should be emphasized that temperature has no effect on the maximum current and the maximal height of the conductance peak, affecting only the line widths in the conductance spectrum.

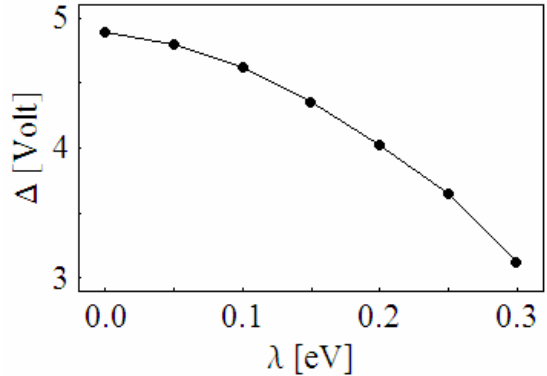

Figure 4: Conductance gap $\Delta$ as a function of vibronic coupling strength $\lambda$. The other parameters of the model are the same as in Fig.2.

Another discrepancy is related to theoretical overestimation of the conductance gap ($\Delta$) due to state-of-art first-principles calculations in comparison with experimental data [38]. Figure 4 presents the $\lambda$-dependence of the $\Delta$-parameter. With an increase of the vibronic coupling strength from 0 to 0.3 eV, the conductance gap is reduced of about 35 % in relation to the initial value. This effect is a direct consequence of the so-called polaron shift in the direction to lower voltages. The approximate formula for the conductance gap has the following form:

$$\Delta \approx 4\left[V_0 - \frac{\lambda^2}{e\Omega} - \frac{4}{\beta} - \Gamma_L - \Gamma_R\right]. \qquad (22)$$

It should be mentioned that the $\Delta$-parameter is mainly determined by the location of the Fermi level with respect to the molecular electronic structure, while it is also influenced by: (i) the strength of the molecule-metal connections, and (ii) the temperature of the system under investigation.

## 4. Summary

Summarizing, we have studied the effect of vibronic coupling on the shape of nonlinear transport characteristics ($I-V$ and $G-V$), using GFT-MT method. This non-perturbative computational scheme is entirely based on mapping which transforms the many-body electron-phonon interaction problem into a one-body multichannel single-electron scattering problem. Here we have shown that inelastic quantum transport through the molecule associated with polaron formation can play a key role in justifying some well-known discrepancies between theoretical calculations and experimental results. However, it should be stressed that the present method on few drastic approximations. For example, here we have completely ignored the following effects: (i) phase decoherence processes in the treatment of the electron-phonon exchange, (ii) Coulomb interactions between charge carriers, and (iii) phonon mediated electron-electron interactions.

Anyway, inelastic transport is quite important for the structural stability [39] and the switching possibility of the molecular electronic devices. Recently, the polaron formation on the molecule was also suggested as a possible mechanism for generating the negative differential resistance (NDR effect) and hysteresis behaviour of the $I-V$ dependence [40]. Moreover, the problem of localized electron-phonon interactions is closely connected to the problem of local heating in current carrying molecular junctions [41,42].

# References

[1] R.C. Jaklevic, and J. Lambe, Phys. Rev. Lett. **17**, 1139 (1966).
[2] C.J. Adkins, and W.A. Phillips, J. Phys. C: Solid State Phys. **18**, 1313 (1985).
[3] K.W. Hipps, and U.J. Mazur, J. Phys. Chem. **97**, 7803 (1993).
[4] B.C. Stipe, M.A. Rezaei, and W. Ho, Science **280**, 1732 (1998).
[5] H.J. Lee, and W. Ho, Science **286**, 1719 (1999).
[6] B.C. Stipe, M.A. Rezaei, and W. Ho, Phys. Rev. Lett. **82**, 1724 (1999).
[7] L.J. Lauhon, and W. Ho, Phys. Rev. B **60**, R8525 (1999).
[8] H. Park, J. Park, A.K.L. Lim, E.H. Anderson A.P. Alivisatos, and P.L. McEuen, Nature **407**, 57 (2000).
[9] J.R. Hahn, H.J. Lee, and W. Ho, Phys. Rev. Lett. **85**, 1914 (2000).
[10] J. Gaudioso, J.L. Laudon, and W. Ho, Phys. Rev. Lett. **85**, 1918 (2000).
[11] L. Lorente, M. Persson, L.J. Lauhon and W. Ho, Phys. Rev. Lett. **86**, 2593 (2001).
[12] J.R. Hahn and W. Ho, Phys. Rev. Lett. **87**, 196102 (2001).
[13] R.H.M. Smit, Y. Noat, C. Untiedt, N.D. Lang, M.C.V. Hemert and J.M.V. Ruitenbeek, Nature **419**, 906 (2002).
[14] N.B. Zhitenev, H. Meng and Z. Bao, Phys. Rev. Lett. **88**, 226801 (2002).
[15] L.H. Yu, Z.K. Keane, J.W. Ciszek, L. Cheng, M.P. Steward, J.M. Tour, and D. Natelson, Phys. Rev. Lett. **93**, 266802 (2004).
[16] J.G. Kushmerick, J. Lazorcik, C.H. Patterson and R. Shashidhar, Nano Lett. **4**, 639 (2004).
[17] W. Wang, T. Lee, I. Kretzschmar and M.A. Reed, Nano Lett. **4**, 643 (2004).
[18] B.N.J. Persson, and A. Baratoff, Phys. Rev. Lett. **59**, 339 (1987).
[19] T. Mii, S.G. Tikhodeev, and H. Ueba, Phys. Rev. B **68**, 205406 (2003).
[20] A. Troisi, M.A. Ratner, and A. Nitzan, J. Chem. Phys. **118**, 6072 (2003).
[21] M. Galperin, M.A. Ratner, and A. Nitzan, Nano Lett. **4**, 1605 (2004).
[22] J. Jiang, M. Kula, W. Lu, and Y. Luo, Nano Lett. **5**, 1551 (2005).
[23] J. Bonča, and S.A. Trugman, Phys. Rev. Lett. **75**, 2566 (1995).
[24] J. Bonča, and S.A. Trugman, Phys. Rev. Lett. **79**, 4874 (1997).
[25] K. Haule, and J. Bonča, Phys. Rev. B **59**, 13087 (1999).
[26] E.G. Emberly, and G. Kirczenow, Phys. Rev. B **61**, 5740 (2000).
[27] L.E.F. Foa Torres, H.M. Pastawski, and S.S. Makler, Phys. Rev. B **64**, 193304 (2001).
[28] H. Ness, S.A. Shevlin, and A.J. Fisher, Phys. Rev. B **63**, 125422 (2001).
[29] H. Ness, and A.J. Fisher, Chem. Phys. **281**, 279 (2002).
[30] K. Walczak, cond-mat/0306174 (submitted).
[31] M. Čížek, M. Thoss, and W. Domcke, Phys. Rev. B **70**, 125406 (2004).

[32] B. Dong, H.L. Cui, X.L. Lei, and N.J.M. Horing, Phys. Rev. B **71**, 045331 (2005).
[33] L.E. Hall, J.R. Reimers, N.S. Hush and K. Silverbrook, J. Chem. Phys. **112**, 1510 (2000).
[34] W. Tian, S. Datta, S. Hong, R.G. Reifenberger, J.I. Henderson and C.P. Kubiak,
    J. Chem. Phys. **109**, 2874 (1998).
[35] V. Mujica, M.A. Ratner and A. Nitzan, Chem. Phys. **281**, 147 (2002).
[36] I. Kinoshita, A. Misu and T. Munakata, J. Chem. Phys. **102**, 2970 (1995).
[37] J.P. Gauyacq, A.G. Borisov and G. Raseev, Surf. Sci. **490**, 99 (2001).
[38] S.T. Pantelides, M. Di Ventra, and N.D. Lang, Physica B **296**, 72 (2001).
[39] Z. Yang, M. Chshiev, M. Zwolak, Y.-C. Chen and M. Di Ventra,
    Phys. Rev. B **71**, 041402 (2005).
[40] M. Galperin, M.A. Ratner and A. Nitzan, Nano Lett. **5**, 125 (2005).
[41] D. Segal, and A. Nitzan, Chem. Phys. **281**, 235 (2002).
[42] Y.-C. Chen, M. Zwolak, and M. Di Ventra, Nano Lett. **3**, 1691 (2003).